# Novel Photovoltaic Phenomenon in Manganite/ZnO Heterostructure

Zhang Jiaqi [1], Huang Keke [1], Yang Xiaotian [2], Si Wenzhe [1], Wu Xiaofeng [1], Du Yanyan [1], Cheng Gang [1], and Feng Shouhua [1]*

1 State Key Laboratory of Inorganic Synthesis and Preparative Chemistry, College of Chemistry, Jilin University, Changchun 130012, P. R. China

2 School of Electronic Information and Engineering, Jilin Institute of Architectural and Civil Engineering，Changchun 130118, P. R. China

*Corresponding author (Email: shfeng@jlu.edu.cn)

## Abstract

In this paper, we report a novel photovoltaic phenomenon in a low cost manganite/ZnO p-n heterojunction grown on ITO glass substrate by pulsed laser depositon (PLD) under relative low growth temperature. The heterostructure ITO/$La_{0.62}Ca_{0.29}K_{0.09}MnO_3$(LCKMO)/ZnO/Al exhibits reproducible rectifying characteristics and consists of four parts: aluminum top electrode, ITO bottom electrode, manganite and ZnO semiconductor films. Moreover, the light current can generate under continuous laser ($\lambda$=325nm) irradiation. In this article, we investigate the influence of manganite and ZnO film thickness on the electrical and photoelectric characteristics of the heterostructure at room temperature. The maximum power conversion efficiency (PCE) is achieved not only when the LCKMO and ZnO layer is thin enough, but also when the full space charge layer is sufficient. We obtain the maximum value(0.0145%) of PCE when the thickness of LCKMO and ZnO layer is 25nm and 150nm, respectively. Under this condition, the open circuit voltage is 0.04V, which can be attributed to internal photoemission.

# 1 Introduction

In the past decade, perovskite manganites, especially hole doped manganites, have attracted intense attention[1,2] of the academic community by displaying fascinating physical behaviors, including intrinsic colossal magnetoresistance (CMR) and metal-insulator (M-I) transition.

Many excellent rectifying behaviors have been observed in manganites and n-type semiconductors based heterostructures[3-5]. In comparison with ordinary semiconductor heterojunction, the manganite based heterojunctions have shown numerous fascinating characteristics such as picosecond photoelectric effect[6], high magnetic sensitivity[7] and positive magnetoresistance[8].

Sheng et al observed change in photovoltage due to external magnetic field in manganite-based heterojunction [9]. Influence of film thickness on the physical properties of manganite-based heterojunctions was also studied [10]. Nevertheless, the earliest works were based on single-crystal substrate, such as Nb: SrTiO$_3$. It was also observed that the high growth temperature condition was always imperative for the epitaxial growth of La$_{0.7}$Ce$_{0.3}$MnO$_3$/Nb:SrTiO$_3$[11], La$_{0.7}$Ca$_{0.3}$MnO$_3$/Nb:SrTiO$_3$[12] and La$_{0.7}$Sr$_{0.3}$MnO$_3$/Nb:SrTiO$_3$[13].

ZnO is a traditional n-type semiconductor with potential application value in the next generation UV light emitting diode and laser diode. Although a large structural disorder exhibits between hexagonal ZnO and orthorhombic manganite layers, the excellent rectifying behaviors[14] and photovoltaic characteristics[15] of manganite/ZnO heterostructure still persist. There is a thin depletion layer at the junction interface due to high carrier concentrations of manganites and ZnO layers. The fabrication of ZnO nanowire/La$_{0.65}$Sr$_{0.35}$MnO$_3$ heterojunction and characterization of its rectifying behavior were reported[16]. Feng et al reported the effects of ZnO film thickness on electrical and magnetoresistance characteristics of $La_{0.8}Sr_{0.2}MnO_3$/ZnO heterostructures [17]. Herein, in the $La_{0.62}Ca_{0.29}K_{0.09}MnO_3$(LCKMO)/ZnO heterostructure on ITO glass substrate, we observe excellent rectifying I-V curve and photoelectric behavior when it is irradiated by continuous laser. We also investigate the influence of film thickness on the electrical and photoelectric characteristics of the junctions.

## 2 Experimental

The photovoltaic device has a structure of ITO/LCKMO/ZnO/Al. The ITO-conducting glass substrate (a sheet resistance of 15 Ω/□) was cleaned separately using acetone, ethanol, and de-ionized water for 15 min before deposition.

The LCKMO target was obtained by conventional ceramic sintering process. The single crystals of $La_{0.62}Ca_{0.29}K_{0.09}MnO3$(LCKMO) were first synthesized by hydrothermal method and then were grounded, pressed, and sintered at 1200℃ in air for 4h to formulate a stoichiometric target. The pure ZnO target was commercially purchased from KJMT.

The LCKMO/ZnO heterostructure was deposited onto the ITO glass substrates by pulsed laser deposition (PLD) with a KrF excimer laser beam (Coherent, λ=248nm) having a fluence of 260mJ. Prior to the deposition, the chamber was vacuumed to a base pressure of $3\times10^{-6}$ Torr. During deposition, the substrate temperature was kept at 300℃ and the distance from substrate to target was ~75mm for both LCKMO and ZnO layer deposition. The optimized oxygen partial pressure of 200 and 150 mTorr was used respectively and the film thickness of LCKMO and ZnO layers was precisely controlled by the number of laser pulses. After deposition, the substrates with films were annealed at the temperature of 300℃ for 0.5h and then naturally cooled to room temperature in oxygen pressure of 400 Torr. Subsequently, the thin films were covered with a homemade mask, then the aluminum (Al) pads with an area of 1mm$^2$ which got Ohmic contact were thermal evaporated on ZnO films as top electrodes.

The polycrystallinity of LCKMO and ZnO films was investigated by an X-ray diffractometer (Ultima IV Protectus) with Cu K radiation. The UV-vis spectrum of LCKMO film was recorded by a Shimadzu UV-2550 spectrophotometer. Cross sectional image of the heterostructure was obtained by a scanning electron microscope (FEI Helios 600). Current density-voltage (J-V) characteristics of the ITO/LCKMO/ZnO/Al p-n junctions were measured using computer-programmed Keithley 2440 source meter. The photoelectric behaviors of ITO/LCKMO/ZnO/Al p-n junctions (~1mm$^2$) were further investigated using a 325nm continuous laser and were measured with the same source meter at the ambient temperature. The light intensity of the laser was measured with a photometer (Coherent Fieldmax II).

## 3 Results and discussion

Figure 1(a) displays a typical XRD spectrum recorded from LCKMO/ZnO thin films grown on ITO glass substrates. The thickness of each layer measured by cross sectional SEM study is shown in Figure 1(b) as follows: $d_{ZnO}$-550nm, $d_{LCKMO}$-100nm, $d_{ITO}$-150nm. In comparison, XRD patterns of ITO glass substrate and ITO glass substrate with only LCKMO deposited are also shown in

Figure 1(a). It can be observed that there is a very strong diffraction peak of the ITO glass substrate near 35.2°, following several other weaker peaks. Most of the other peaks can be attributed to ZnO. No diffraction peak of LCKMO thin films was clearly observed. To further clarify this observation, we also conducted the experiment to show that the cross sectional SEM images of ITO/LCKMO/ZnO films in Figure 1(b). LCKMO layer with a thickness of 120nm can be clearly seen. Therefore, it is concluded that the LCKMO thin films deposited on ITO glass substrate under 300℃ are amorphous. Similar results have been observed for several other oxide films prepared under similar deposition temperature [18,19].

Figure 2 shows the J-V characteristics of heterojunctions (ITO/LCKMO/ZnO/Al) with and without continuous laser irradiation and the inset shows the schematic structure of the device. The incident laser is irradiated from the ITO side (bottom). The depletion layer is formed when the LCKMO contacts with ZnO. The holes of LCKMO and the electrons of ZnO near the interface are exhausted. The diffusion barrier is then built up to prevent further movement of the carriers. The heterojunctions maintain high-quality rectifying characteristics during the voltage sweeping from -1V to +1V. It can be clearly observed that the current increases marginally under continuous laser irradiation, both forward and backward within the bias region. One possible mechanism of the current enhancement is that the nonequilibrium carriers are excited by the laser, thus increasing the density of carriers. Another possible mechanism would be the thermal effect.

Figure 3 shows an enlarged view of a part of the J-V curve (from -0.2V to +0.2V). The intensity of the laser beam is 550mW/cm$^2$. The effective area of the devices is 1mm$^2$. The devices show a short circuit current density (Jsc) of 2.3mA/cm$^2$, open circuit voltage (Voc) of 0.04V, fill factor (FF) of 25.69%, and power conversion efficiency (PCE) of 0.0047 %. Upon analyzing the result, it is inferred that the weakness of PCE may be caused by the following reasons. The photocurrent density is in direct proportion to light absorption of the active layer. The optical absorption property of LCKMO films is investigated at room temperature. The UV-vis absorption spectrum is shown in the inset of Figure 3. According to the spectrum, LCKMO has high absorbance from 350 to 400nm but a bit low at 325nm which is the wavelength of the continuous laser. In addition, an amorphous growth of LCKMO films is inevitable for low-cost which has been confirmed. It causes recombination of the nonequilibrium charge carriers in grain boundary. When the laser with the wavelength of 325nm irradiates the heterojunctions, the light-induced nonequilibrium holes and electrons in LCKMO and ZnO are created. Although the photon energy of about 3.8eV is larger than the band gaps of LCKMO and ZnO, the results depict very low open circuit voltage(0.04V) which can be attributed to internal photoemission [12]. An schematic energy diagram is shown in the inset of Figure 5.

We also investigate the influence of film thickness of the junctions which have structures of ITO/LCKMO/ZnO/Al. As well known, the film thickness can affect the electric properties by influencing depletion width and interfacial barrier. For the photoelectric process, it may affect the diffusion and number of the nonequilibrium charge carriers. Eventually it will have an effect on the power conversion efficiency of the heterostructures [10]. The thickness of ZnO(n type) films is firstly fixed as 550nm and we change the thickness of LCKMO layers. Figure 4(a) shows the variation of open circuit voltage ($V_{oc}$) and power conversion efficiency (PCE) when the thickness (d) of LCKMO(p type) films has been changed. Detailed results are provided in the inset of Figure 4(a). When d=100nm, $V_{oc}$ is 0.04V and PCE is 0.0014%. The thickness of the LCKMO layer is sufficient to obtain full space charge layer. When d=50nm, $V_{oc}$ is 0.04V and PCE is 0.0047%. PCE is larger than that of 120nm for smaller resistance caused by thinner films. When d=25nm, $V_{oc}$ is 0.04V and PCE is 0.0097% as maximum value. When d=15nm, $V_{oc}$ is 0.02V and PCE is 0.0069%, the reduction of $V_{oc}$ indicates LCKMO layer is fully depleted and the thickness of space charge layer decreases. When d=10nm, $V_{oc}$ is 0.01V and PCE is 0.0025%. The thickness of space charge layer further decreases. When d=7.5nm, $V_{oc}$ and PCE are both 0. Because surface roughness of commercial ITO glass substrate is approximately 6nm(measured by atomic force microscope, not shown here), the layer is so thin that we cannot prevent the contact of ZnO and bottom electrode. Another possible reason may be tunneling transport of the charge carriers. The maximum of PCE is achieved when not only LCKMO layer is thin enough but also when the full space charge layer is confirmed.

Then we change the film thickness of ZnO. The thickness of LCKMO films is fixed as 25nm. When d is 550nm, $V_{oc}$ is 0.04V and PCE is 0.0097%. When d decreases to 250nm, $V_{oc}$ is 0.04V and PCE increases up to 0.0099%. When d is 150nm, $V_{oc}$ is 0.04V and PCE is 0.0145% as maximum value. When d is 50nm, Voc and PCE are both 0. It can be observed that the influence of ZnO films thickness has the same feature as that of LCKMO. As well, we obtain the maximum value (0.0145%) of PCE when the thickness of LCKMO and ZnO is 25nm and 150nm respectively.

The current study of heterojunctions can be attributed to one of the Schottky (metal-semiconductor) barrier type, having the form $I = I_0(\exp(eV/\eta kT) - 1)$, where $I_0$ is the saturation current and ŋ is the ideality factor. Fitting results show $I_0$ of $7.2 \times 10^{-5}$ A and ŋ of 18 [Figure 5]. As per results of the previous discussion, we surmise that the suitable transport mechanism is the thermionic emission model which is valid for the Schottky barriers, when the

$$I_0 = R^* T^2 \exp(-\phi_B / V_T)$$

interface barrier presents a substantially important impediment to the current flow, having the form                         .

Further studies on the enhancement of power conversion efficiency and the modification of rectifying behavior are underway.

## 4 Conclusion

In conclusion, a low-cost photovoltaic application of manganite-based heterojunctions having a form ITO/LCKMO/ZnO/Al is studied. The device exhibits reproducible rectifying characteristics. Photocurrent is observed under continuous laser irradiation. The exploration of low-cost application includes four aspects: aluminum as top electrodes, ITO as bottom electrodes, ZnO as n-type semiconductor and relative low growth temperature. Our results suggest the possibility to develop low-cost manganite-based photovoltaic devices. We also independently investigated the influence of manganite and ZnO film thickness on the electrical and photoelectric characteristics of the junctions. The maximum of PCE is achieved when not only LCKMO and ZnO layers are thin enough but also full space charge layer is confirmed. We get the maximum value(0.0145%) of PCE when the thickness of LCKMO and ZnO is 25nm and 150nm, respectively. Open circuit voltage of the devices is 0.04V which can be attributed to internal photoemission.

## Acknowledgments

This work was supported by the National Natural Science Foundation of China (No.90922034 and No.21131002) and Specialized Research Fund for the Doctoral Program of Higher Education (No. 20110061130005).

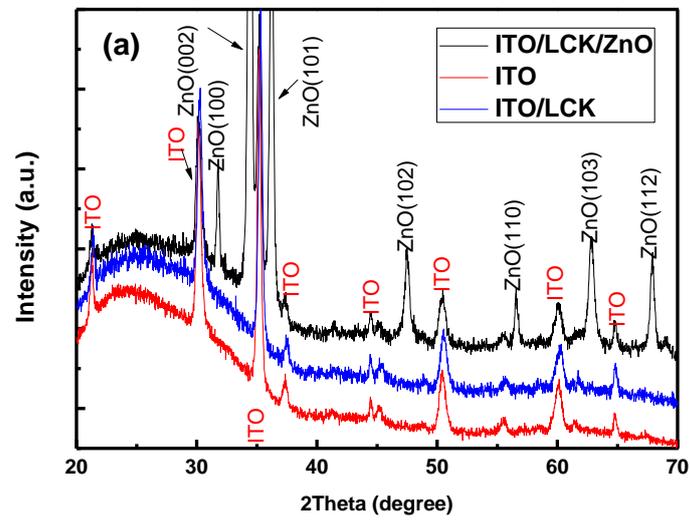

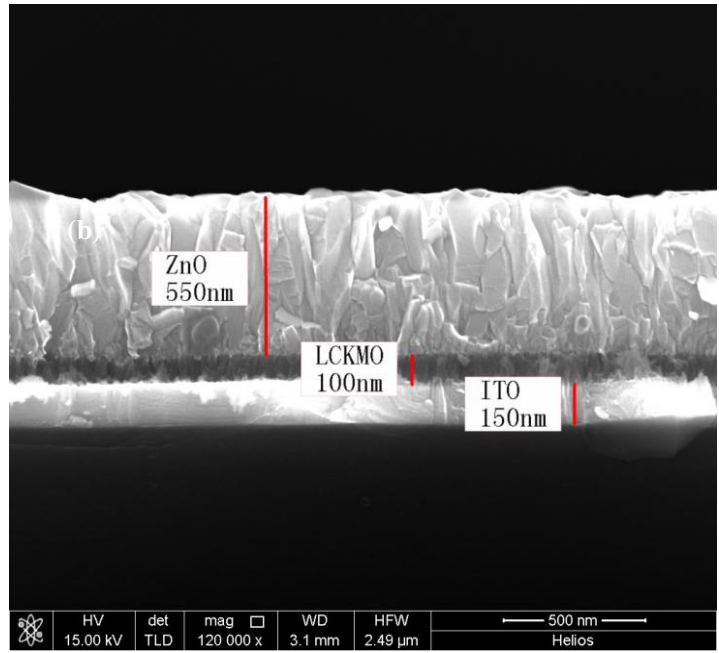

Fig.1. (a) XRD patterns recorded from ITO, ITO/LCKMO and ITO/LCKMO/ZnO samples. (b) Cross-sectional SEM image of the ITO/LCKMO/ZnO structures.

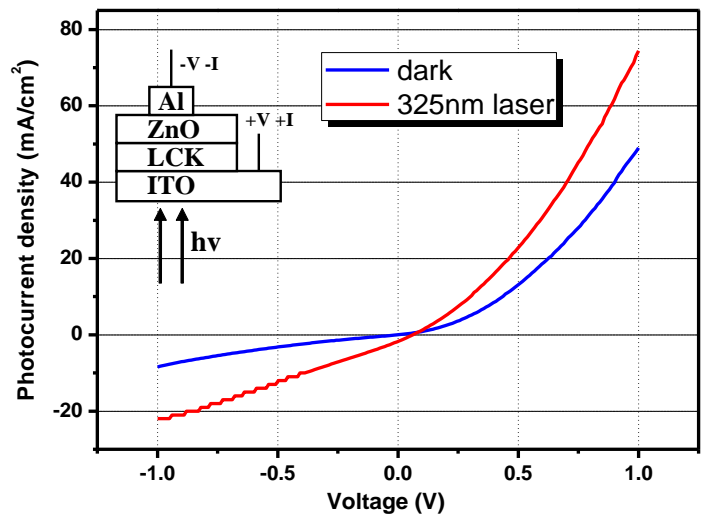

Fig.2. J-V characteristics of device ITO/LCKMO/ZnO/Al with and without continuous laser illumination. Inset: the schematic structure drawing of the devices and method of measurement.

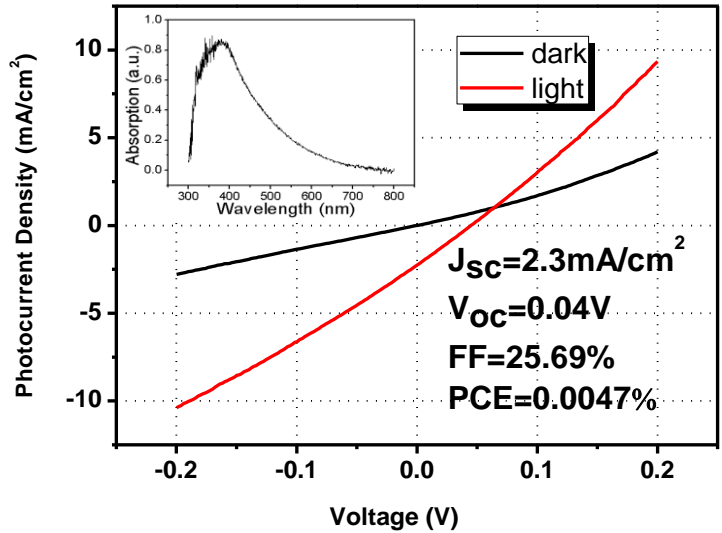

Fig.3. The enlarged view of the part of the J-V curves. Inset: UV-visible spectra of LCKMO films.

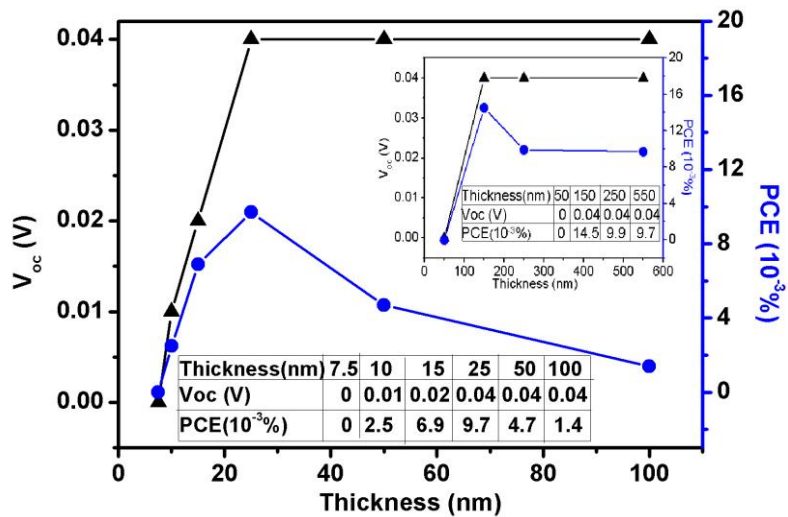

Fig.4. Open circuit voltage ($V_{oc}$) and power conversion efficiency (PCE) as functions of LCKMO and ZnO (top inset) film thickness. Detailed results are given in the bottom inset.

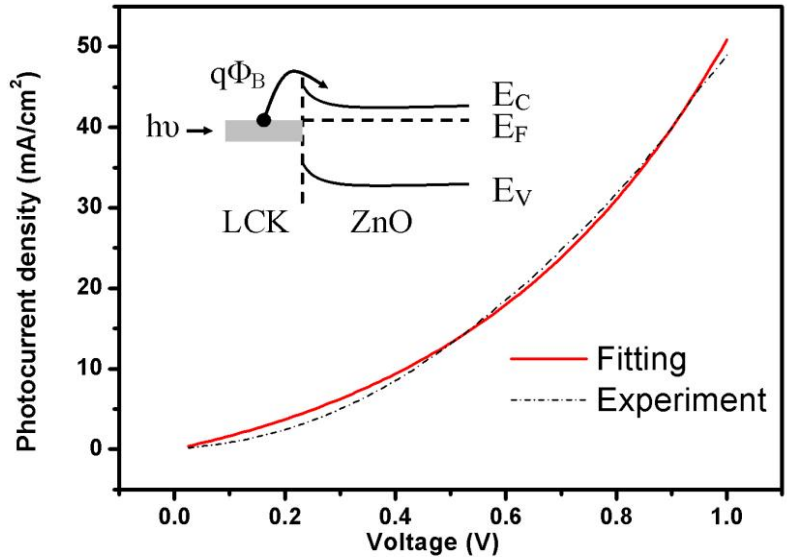

Fig.5. J-V curves corresponding to fitting and experiment results. Inset: The schematic energy diagram of the heterojunctions.